\documentclass[12pt]{article}
\usepackage{amsmath}
\usepackage{amsfonts}
\usepackage{amssymb}
\usepackage{cancel}
\usepackage{indentfirst}
\usepackage{bbold}
\usepackage{enumerate}
\usepackage{hyperref}
\usepackage{cite}
\usepackage{authblk}
\title{\vskip-2in\hbox to\textwidth{\hfill \normalsize YITP-SB-13-6}\vskip1.8in
Systematizing semi-shortening}
\author{C.-Y. Ju\thanks{\href{mailto:cju@insti.physics.sunysb.edu}{cju@insti.physics.sunysb.edu}} and W. Siegel\thanks{\href{mailto:siegel@insti.physics.sunysb.edu}{siegel@insti.physics.sunysb.edu}}}
\affil{\it C. N. Yang Institute for Theoretical Physics\\
State University of New York, Stony Brook, NY 11794-3840}
\date{}
\begin{document}
	\maketitle
	\begin{abstract}
		We re-derive semi-shortening conditions for four-dimensional superconformal field theory with a different approach. These conditions have similar patterns that can be generalized to weaker constraints, including all those of F. Dolan and H. Osborn.  In particular, for the case of $\mathcal{N} = 4$ super Yang-Mills theory, formulated in projective superspace, we find constraints for all BPS operators. We also give an example how constraints can be found from known ones. These constraints are a subset of our maximal set of semi-shortening conditions.
	\end{abstract}
	
	\section{Introduction}
	
	\numberwithin{equation}{section}
	
	The AdS/CFT correspondence\cite{Maldacena:1997re} provides a good way of understanding $\mathcal{N}=4$ super Yang-Mills theories or supergravity qualitatively\cite{Beisert:2010jr}. 	
	For a superconformal theory to be a valid quantum theory, it has to satisfy some unitarity bounds\cite{Dobrev:1985qv,Minwalla:1997ka}. When the bound is saturated, i.e., when the inequality becomes equality, the primary field loses some degrees of freedom. This implies the primary state can be annihilated by some combination of super charge and vice versa (Bogomol'nyi-Prasad-Sommerfield conditions). A supermultiplet satisfying a BPS condition will be truncated into a shorter supermultiplet\cite{Andrianopoli:1998jh,Andrianopoli:1998ut,Ferrara:1999ed}, hence it is also called a shortening condition. Various short and semi-short representations for $\mathcal{N}=2$ and $\mathcal{N}=4$ in four dimension are discussed in \cite{Dolan:2002zh}.
	
	In this paper, we first review how shortening conditions can be treated as defining coset superspaces\cite{Heslop:2000af}. We then show how most semi-shortening conditions in four dimensions can be obtained by superconformally transforming the massless field equation. The remaining known (semi-)shortening conditions can then be obtained by a simple generalization. Finally, we consider the example of $\mathcal{N}=4$ SYM and apply the algorithm to find explicit expression for semi-shortening constraints.
	
	\section{Coset superspace\label{coset}}	
	
	In ordinary quantum mechanics, wave functions are defined as 
	\begin{equation*}
		\phi (x) = \left< x | \phi \right>.
	\end{equation*}
	$\left<x\right|$ in the above equation is the coordinate basis of Hilbert space, which can be written as
	\begin{equation*}
		\left< x \right| \equiv \left< 0 \right| U^{-1}(x) = \left< 0 \right| e^{-ix\hat{P}},
	\end{equation*}
	where $\left| 0 \right>$ is the ground state or the highest weight state.
	
	In supersymmetric field theories, we can generalize the ``propagator" from $U(x) = e^{ix\hat{P}}$ to $U(x,\theta) = e^{i\left(x\hat{P} + \theta \hat{Q} + \bar{\theta} \hat{\bar{Q}}\right)}$, a general element of the supersymmetry group. The coordinate basis in the Hilbert space becomes $\left< x,\theta \right| = \left< 0 \right| U^{-1}(x,\theta,\bar{\theta})$. Then the ``field'' of an arbitrary state in the Hilbert space $\left| \phi \right>$ is again $\phi (x,\theta) \equiv \left< x,\theta\right|\left.\phi\right>$, which is called a {\it superfield}. This formulation not only includes all the superpartner fields automatically but also gives supersymmetry theory an interesting geometric meaning: Supersymmetry can be treated as adding extra Grassmann coordinates to the ordinary spacetime coordinates. This generalized space is called {\it superspace}. (More details can be found in \cite{Gates:1983nr}). 
	
	With this superspace formulation, the generators can be written as derivatives. As the simplest example, the generator of spacetime coordinate translations can be written as the derivative $P_\mu = i\frac{\partial}{\partial x^\mu}$. We can check by acting $P_\mu$ on an arbitrary superfield $\phi(z)$, in which $z = (x,\theta,\bar{\theta})$:
	\begin{eqnarray*}
		P_\mu \phi(z) & = & i\left< 0 \right| \frac{\partial}{\partial x^\mu} e^{-i(x\hat{P} + \theta \hat{Q} + \bar{\theta} \hat{\bar{Q}})} \left| \phi \right>\\
		& = & \left< 0 \right| e^{-i(x\hat{P} + \theta \hat{Q} + \bar{\theta} \hat{\bar{Q}})} \hat{P}_\mu \left| \phi \right>.
	\end{eqnarray*}
	By the same token, we can also construct other generators as derivatives. Not all the generators commute with the propagator, therefore, the ordering matters. We define symmetry generators, $G$, on superfields as
	\begin{equation*}
		G\phi(z) = \left< 0 \right| U^{-1}(z) \hat{G} \left| \phi \right>,
	\end{equation*}
	and the covariant derivatives, $g$, as
	\begin{equation*}
		g \phi(z) = \left< 0 \right|  \left(-\hat{G}\right) U^{-1}(z) \left| \phi \right>,
	\end{equation*}
	where $\hat{G}$ can be any generator of superspace\cite{Siegel:2010yd}. 
Thus symmetry and covariant derivatives correspond to right and left multiplication, respectively, on the group element represented by the propagator.  Since left and right multiplication commute, the covariant derivatives are actually {\it invariant} under symmetry transformations.  (This invariance becomes only covariance if the coset constraints are used to fix a ``gauge" where some of the constrained coordinates vanish: See below.)
In the usual supersymmetry theory, the generators are $\lbrace \hat P, \hat Q, \hat{\bar{Q}}, \hat M, \hat{\bar{M}}, \hat R \rbrace$ which correspond to translation, supersymmetry, rotation, and $R$-symmetry.
	
	In superconformal field theory, in addition to the usual generators, there are also $\lbrace \hat K,\hat S, \hat{\bar{S}}, \hat D \rbrace$, known as the generators of special conformal transformations, superconformal transformations, and dilatation. In $D=4$, the superconformal group is $(P)SU(2,2|\mathcal{N})$. We can wick rotate to $(P)SL(4|\mathcal{N}$) and treat not only ``projective'' ($P$) but also ``special'' ($S$) as gauge invariances. Then the group before gauge fixing is $GL(4|\mathcal{N})$. The coordinates of the full superspace, $z_{\mathcal{ M }}^{~~\mathcal{ A }}$, can be ordered as follows
	\begin{eqnarray}
		z_\mathcal{M}^{~~\mathcal{A}} & = & \bordermatrix{
			~ & \alpha & i & \dot{\alpha}\cr
			\beta & z_{\beta}^{~\alpha} & z_{\beta}^{~i} & z_{\beta}^{~\dot{\alpha}} \cr
			j & z_{j}^{~\alpha} & z_{j}^{~i} & z_{j}^{~\dot{\alpha}} \cr
			\dot{\beta} & z_{\dot{\beta}}^{~\alpha} & z_{\dot{\beta}}^{~i} & z_{\dot{\beta}}^{~\dot{\alpha}}
		}\nonumber\\
		& = & \begin{pmatrix}
			\text{\footnotesize Lorentz$+$scale} & \text{\footnotesize supersymmetry} & \text{\footnotesize translation}\\
			\text{\footnotesize superconformal} & \text{\footnotesize R-symmetry} & \text{\footnotesize supersymmetry}\\
			\text{\footnotesize special conformal} & \text{\footnotesize superconformal} & \text{\footnotesize Lorentz$-$scale}
		\end{pmatrix}.\label{coordinate}
	\end{eqnarray}
	Throughout this paper, all the Greek indices are spinor (fermionic) indices, Latin indices stand for internal/R-symmetry (bosonic) indices, and calligraphic capital Latin indices can be both.
	The full superspace propagator can be written as $U(z) = \exp\left(iz\hat{G}\right)$, where $\hat{G}$ is the corresponding symmetry generator. If the ground state is invariant under some symmetries (with corresponding symmetry generators $\hat{H}_\iota$), we can divide symmetry generators into two groups, $\hat{G} = \lbrace \hat{T}_i , \hat{H}_\iota \rbrace$. Then the ground state propagates as
	\begin{eqnarray*}
		& & \exp\left(iz\hat{G}\right) \left| 0 \right> = \exp\left(i\tilde{z}\hat{T}\right) \exp\left(i\zeta\hat{H}\right)\left| 0 \right> = \exp\left(i\tilde{z}\hat{T}\right) \left| 0 \right>\\
		& & \Rightarrow \tilde{U}(\tilde{z}) = \exp\left(i\tilde{z}\hat{T}\right) = \exp\left(iz\hat{G}\right) ~mod~\hat{H}_\iota = U(z) ~mod~ \hat{H}_\iota.
	\end{eqnarray*}
	In other words, the full superspace becomes a coset superspace. Therefore, we can set the coordinates corresponding to $H$ to zero. For example, to get the usual superspace, we gauge away the lower-left triangle and the diagonal parts of the coordinate matrix as 
	\begin{eqnarray*}
		z_\mathcal{M}^{~~\mathcal{A}} & \rightarrow & \tilde{z}_\mathcal{M}^{~~\mathcal{A}} = \begin{pmatrix}
		\mathbb{1} & \theta_{\alpha'}^{~i} & x_{\alpha'}^{~\dot{\alpha}} \cr
		0 & \mathbb{1} & \bar{\theta}_{i'}^{~\dot{\alpha}} \cr
		0 & 0 & \mathbb{1}
		\end{pmatrix}.
	\end{eqnarray*}

	We can also treat projective superspaces as coset superspaces by modding out some coordinates.  Rearranging the full coordinate matrix as
	\begin{eqnarray*}
		z_\mathcal{M}^{~~\mathcal{A}} = \bordermatrix{
			~ & \alpha & i & i' & \dot{\alpha}\cr
			\beta & z_{\beta}^{~\alpha} & z_{\beta}^{~i} & z_{\beta}^{~i'} & z_{\beta}^{~\dot{\alpha}} \cr
			j & z_{j}^{~\alpha} & z_{j}^{~i} & z_{j}^{~i'} & z_{j}^{~\dot{\alpha}} \cr
			j' & z_{j'}^{~\alpha} & z_{j'}^{~i} & z_{j'}^{~i'} & z_{j'}^{~\dot{\alpha}}\cr
			\dot{\beta} & z_{\dot\beta}^{~\alpha} & z_{\dot\beta}^{~i} & z_{\dot\beta}^{~i'} & z_{\dot\beta}^{~\dot{\alpha}}
		},
	\end{eqnarray*}
	($i$ runs from 1 to $n$ and $i'$ from $n+1$ to $\mathcal{N}$), we again gauge away the lower-left and the diagonal blocks
	\begin{eqnarray}
		z_\mathcal{M}^{~~\mathcal{A}} \rightarrow \begin{pmatrix}
			\mathbb{1} & 0 & \theta_{\beta}^{~i'} & x_{\beta}^{~\dot{\alpha}} \cr
			0 & \mathbb{1} & u_{j}^{~i'} & \bar{\theta}_{j}^{~\dot{\alpha}} \cr
			0 & 0 & \mathbb{1} & 0\cr
			0 & 0 & 0 & \mathbb{1}
		\end{pmatrix}.\label{procoord}
	\end{eqnarray}
	This is a consequence of setting the ground state to be annihilated by $\hat{Q}^i$'s and $\hat{\bar{Q}}^{i'}$'s in full superspace.

	Take $\mathcal{N}=4$ SYM field strength in projective superspace as an example, it can be expanded into component fields as follows:
		
	\begin{align*}
		\varphi = & (\phi + u_i^{~i'}\phi_{i'}^{~i} + \frac{1}{2} u^2 \bar{\phi}) + \theta_\alpha^{~i'} (\lambda_{i'}^{~\alpha}+u_{~i'}^{i}\lambda_{i}^{\alpha})+\bar{\theta}_i^{~\dot{\alpha}}(\bar{\lambda}_{\dot{\alpha}}^{~i}+u_{~i'}^{i}\bar{\lambda}_{\dot{\alpha}}^{~i'})\\
		& + \theta^2_{\alpha\beta}f^{\alpha\beta} + \bar{\theta}^{2\dot{\alpha}\dot{\beta}}\bar{f}_{\dot{\alpha}\dot{\beta}} - i \theta_{\alpha}^{~i'}\bar{\theta}_{i}^{\dot{\alpha}}\partial_{\dot{\alpha}}^{~\alpha}(\phi_{i'}^{~i}+u_{~i'}^{i}\bar{\phi})-i\theta^2_{\alpha\beta}\bar{\theta}_i^{~\dot{\alpha}}\partial_{\dot{\alpha}}^{~\alpha}\lambda^{i\beta}\\
		& - i \bar{\theta}^{2\dot{\alpha}\dot{\beta}}\theta_\alpha^{~i'}\partial_{\dot{\alpha}}^{~\alpha}\bar{\lambda}_{\dot{\beta}i'} - \theta^2_{\alpha\beta}\bar{\theta}^{2\dot{\alpha}\dot{\beta}}\partial_{\dot{\alpha}}^{~\alpha}\partial_{\dot{\beta}}^{~\beta}\bar{\phi},
	\end{align*}
	where we have used the internal $SL(2)^2$ to raise and lower the indices. The ``incomplete'' expansion of $u$'s and $\theta$'s is explained at the beginning of subsection \ref{N=4} (equation \ref{rr}).
	
	\section{Shortening conditions as coset space\label{short}}
	
	As in section \ref{coset}, the covariant derivatives for superconformal symmetry can be written as the following graded matrix:
	\begin{eqnarray}
		g_\mathcal{M}^{~~\mathcal{N}} & = & \bordermatrix{
			~  & \alpha & i & \dot{\alpha}\cr
			\beta & g_{\beta}^{~\alpha} & g_{\beta}^{~i} & g_{\beta}^{~\dot{\alpha}} \cr
			j & g_{j}^{~\alpha} & g_{j}^{~i} & g_{j}^{~\dot{\alpha}} \cr
			\dot{\beta} & g_{\dot\beta}^{~\alpha} & g_{\dot\beta}^{~i} & g_{\dot\beta}^{~\dot{\alpha}}
		}\nonumber\\
		& = & \bordermatrix{
			~  & \alpha & i & \dot{\alpha}\cr
			\beta & m_{\beta}^{~\alpha} - i \frac{1}{2}\delta_{\beta}^{~\alpha} d & s_{\beta}^{~i} & k_{\beta}^{~\dot{\alpha}} \cr
			j & q_{j}^{~\alpha} & r_{j}^{~i} & \bar{s}_{j}^{~\dot{\alpha}} \cr
			\dot{\beta} & p_{\dot\beta}^{~\alpha} & \bar{q}_{\dot\beta}^{~i} & m_{\dot\beta}^{~\dot{\alpha}} + i\frac{1}{2} \delta_{\dot\beta}^{~\dot{\alpha}} d
		}\label{matrix}\\
		& = & \begin{pmatrix}
			\text{\footnotesize Lorentz$+$scale} & \text{\footnotesize superconformal} & \text{\footnotesize special conformal}\\
			\text{\footnotesize supersymmetry} & \text{\footnotesize R-symmetry} & \text{\footnotesize superconformal}\\
			\text{\footnotesize translation} & \text{\footnotesize supersymmetry} & \text{\footnotesize Lorentz$-$scale}
		\end{pmatrix}\nonumber
	\end{eqnarray}
	
	In our conventions, the (anti)commutation relations are
	\begin{eqnarray}
		\left[ g_\mathcal{M}^{~~\mathcal{N}}, g_\mathcal{P}^{~~\mathcal{Q}}\right\rbrace = \delta^\mathcal{N}_\mathcal{P}g_\mathcal{M}^{~~\mathcal{Q}} - \left( -1 \right)^{(\mathcal{M}+\mathcal{N})(\mathcal{P}+\mathcal{Q})} \delta^\mathcal{Q}_\mathcal{M}g_\mathcal{P}^{~~\mathcal{N}}, \label{CR}
	\end{eqnarray}
	where in the exponent of $-1$
	\begin{eqnarray*}
		\mathcal{A} = \left\lbrace
		\begin{array}{lcl}
			0 & , & \mathcal{A} \in \text{bosonic} \\
			1 & , & \mathcal{A} \in \text{fermionic}
		\end{array}\right. .
	\end{eqnarray*}
	
	The usual shortening conditions restrict some $g_i^{~\alpha} \phi = 0$ or $g_{\dot{\alpha}}^{~i} \phi = 0$ (``antichiral'' or ``chiral''). Together with superconformal symmetry, the shortening conditions imply the superfield also vanishes under some R-symmetry charges or Lorentz $\pm$ scale generators, by closure of the algebra. We can, therefore, set the left-bottom of the coordinate matrix (special conformal and superconformal coordinates) and some blocks at the right-top (``chiral'' or ``antichiral'' invariant and the symmetries induced) to zero.
	
	It is worth mentioning that the shortening conditions obtained from $g_i^{~\alpha} \phi = 0$ (i.e., $q_i^{~\alpha} \phi = 0$) form a closed set (as do $g_{\dot{\alpha}} ^{~i} \phi = 0$) that doesn't include other $g_j^{~\beta}$ or $g_{\dot{\beta}}^{~j}$. Derivation details are in appendix \ref{q}.
	
	Take projective superspace as an example: We first divide R-symmetry indices into two categories ($i$, $i'$). Some superspace coordinates vanish under some supercharges, $g_i^{~\alpha}\phi = 0$ and $g_{\dot{\alpha}}^{~i'} \phi = 0$. These conditions set some R-symmetry charges acting on the superfield to vanish (see appendix \ref{q}). Therefore this  gives the coordinate matrix shown in equation (\ref{procoord}).
	
	We then consider the general case of superspaces with chiral, antichiral, or ``achiral" fermionic coordinates. R-symmetry indices can be split into three parts ($i, i', i''$), where $i$ is antichiral, $i'$ is achiral, and $i''$ is chiral. Then the generator matrix can be written as follows:
	\begin{eqnarray*}
		g_\mathcal{M}^{~~\mathcal{N}} & = & \bordermatrix{
			~  & \alpha & i & i' & i'' & \dot{\alpha}\cr
			\beta & g_{\beta}^{~\alpha} & g_{\beta}^{~i} & g_{\beta}^{~i'} & g_{\beta}^{~i''} & g_{\beta}^{~\dot{\alpha}} \cr
			j & g_{j}^{~\alpha} & g_{j}^{~i} & g_{j}^{~i'} & g_{j}^{~i''} & g_{j}^{~\dot{\alpha}} \cr
 			j' & g_{j'}^{~\alpha} & g_{j'}^{~i} & g_{j'}^{~i'} & g_{j'}^{~i''} & g_{j'}^{~\dot{\alpha}} \cr
			j'' & g_{j''}^{~\alpha} & g_{j''}^{~i} & g_{j''}^{~i'} & g_{j''}^{~i''} & g_{j''}^{~\dot{\alpha}} \cr
			\dot{\beta} & g_{\dot\beta}^{~\alpha} & g_{\dot\beta}^{~i} & g_{\dot\beta}^{~i'} & g_{\dot\beta}^{~i''} & g_{\dot\beta}^{~\dot{\alpha}}
		}\\
		& = & \bordermatrix{
			~  & \alpha & i & i' & i'' & \dot{\alpha}\cr
			\beta & \times & \times & \times & \times & \times \cr
			j & \times & \otimes & \otimes & \otimes & \times \cr
 			j' & q_{j'}^{~\alpha} & r_{j'}^{~i} & r_{j'}^{~i'} & \otimes & \times \cr
			j'' & q_{j''}^{~\alpha} & r_{j''}^{~i} & r_{j''}^{~i'} & \otimes & \times\cr
			\dot{\beta} & p_{\dot\beta}^{~\alpha} & \bar{q}_{\dot\beta}^{~i} & \bar{q}_{\dot\beta}^{~i'} & \times &\times
			},
	\end{eqnarray*}
	where ``$\times$'' mean it is zero by construction, ``$\otimes$'' is ``induced'' to zero.	And therefore, the gauged coordinate matrix is
	\begin{eqnarray*}
		z_\mathcal{M}^{~~\mathcal{N}} & = & \bordermatrix{
			~  & \alpha & i & i' & i'' & \dot{\alpha}\cr
			\beta & \mathbb{1} & 0 & \theta_{\beta}^{~i'} & \theta_{\beta}^{~i''} & x_{\beta}^{~\dot{\alpha}} \cr
			j & 0 & \mathbb{1} & u_{j}^{~i'} & u_{j}^{~i''} & \bar{\theta}_{j}^{~\dot{\alpha}} \cr
 			j' & 0 & 0 & u_{j'}^{~i'} & u_{j'}^{~i''} & \bar{\theta}_{j'}^{~\dot{\alpha}} \cr
			j'' & 0 & 0 & 0 & \mathbb{1} & 0 \cr
			\dot{\beta} & 0 & 0 & 0 & 0 & \mathbb{1}
		}
	\end{eqnarray*}

	\section{On-shell constraints\label{onshell}}
		
	By definition, superconformal primary superfields must satisfy the conditions
		\begin{eqnarray*}
			s_\alpha^{~i} \phi (z) = 0 ~ \text{and} ~ \bar{s}_{i}^{~\dot{\alpha}}\phi (z) = 0,
		\end{eqnarray*}
		which also implies $k_{\dot{\alpha}}^{~\alpha} \phi(z) = 0$. Note that these are covariant derivatives, not symmetry generators.
		
		For a massless free field, the superfield has to satisfy the on-shell condition $p^2 \phi (z)=0$. However, this condition is not invariant under superconformal transformations. (These are not symmetry transformations, except on the vacuum.  On the superfield, they are transformations generated by the coset constraints.) This can easily be seen from the following example:
		\begin{eqnarray}
			0 & = &p^2 \phi\nonumber\\
			\Rightarrow 0 & =  & s_\alpha^{~i} p^2 \phi\nonumber\\
			& = & \left(\left[ s_\alpha^{~i} , p^2 \right]  + p^2 s_i^{~\alpha}\right) \phi\nonumber\\
			& = & \left[ s_\alpha^{~i} , p^2 \right] \phi\nonumber\\
			\Rightarrow 0 & = & p^{\dot{\alpha}}_{~\alpha}\bar{q}_{\dot{\alpha}}^{~i}\phi.\label{p^2}
		\end{eqnarray}
		Therefore, a superconformal, massless, free field should also satisfy constraint eq.\ (\ref{p^2}). One can keep applying $s$ or $\bar{s}$ to get more constraints on the massless superfield\cite{Siegel:1987ik}. Since both $s$ and $\bar{s}$ are fermionic operators, the number of constraints on the field is finite. The constraints can be represented diagrammatically as follows:
		\begin{eqnarray*}
				&& ~~~~~~~~~~~~~~~~~~~~~~~~~~~~\lbrace 0 \rbrace \ p^2\phi = 0\\
				&& ~~~~~~~~~~~~~~~~~~~~~~~~~~~~~~\swarrow ~~~~~ {\searrow}\\
				&& ~~~~~~~~~~~~~~~~~~~~~~~~s\swarrow ~~~~~~~~~~~~~~~ {\searrow}~ \bar{s}\\
				&& ~~~~~~~~~~~~~~~~~~~~~\swarrow ~~~~~~~~~~~~~~~~~~~~~~~~~ { \searrow}\\
				&& ~~~~~~~~~~~~~~~~ \lbrace 1\rbrace ~~~~~~~~~~~~~~~~~~~~~~~~~~~~~~~~~~~ \lbrace 2\rbrace\\
				&& ~~~~~~~~~~ \swarrow ~~~~~~~ \searrow ~~~~~~~~~~~~~~~~~~~~~~~\swarrow ~~~~~~~ \searrow\\
				&& ~~~~~~ \lbrace 3\rbrace ~~~~~~~~~~~~~~~ \lbrace 4\rbrace ~~~~~~~~~~~~~~~~~~ \lbrace 5\rbrace ~ ~~~~~~~~~~~~~~ \lbrace 6\rbrace\\
				&& ~~\swarrow ~ \searrow  ~~~~~~~~~~\swarrow  ~ \searrow ~~~~~~~~~~~\swarrow ~~~ \searrow ~~~~~~~~~~ \swarrow ~~~ \searrow \\
				&& {\cancelto{0}{\lbrace 7\rbrace}} ~~~~~~ \lbrace 8\rbrace ~~~~~~\lbrace 9\rbrace ~~~~~ \lbrace 10\rbrace ~~~~~~\lbrace 11\rbrace ~~~~~~ \lbrace 12\rbrace ~~~~~ \lbrace 13\rbrace ~~~~~~ {\cancelto{0}{\lbrace 14\rbrace}}\\
				&& ~~~~~~~~~\swarrow  \searrow ~~~\swarrow \searrow ~~~\swarrow \searrow ~~~~~\swarrow \searrow ~~~~\swarrow \searrow ~~~~~\swarrow \searrow\\
				&&~~~~~~ {\cancelto{0}{\lbrace 15\rbrace}} \lbrace 16\rbrace {\cancelto{0}{\lbrace 17\rbrace}} \lbrace 18\rbrace \lbrace 19\rbrace {\cancelto{0}{\lbrace 20\rbrace}} {\cancelto{0}{\lbrace 21\rbrace}} \lbrace 22\rbrace \lbrace 23\rbrace {\cancelto{0}{\lbrace 24\rbrace}} \lbrace 25\rbrace {\cancelto{0}{\lbrace 26\rbrace}}
		\end{eqnarray*}\\
	$\cancelto{0}{\lbrace \# \rbrace}$\quad in the diagram means it is identically zero by the coset constraints $s=\bar s=k=0$, hence doesn't imply any new constraints. All the semi-shortening conditions in the diagram are compatible with $p^2 = 0$. The full constraints obtained from $p^2=0$ are listed in appendix \ref{App}. It is worth mentioning that this formalism is very general in that it automatically includes {\it all} semi-shortening conditions quadratic in covariant derivatives:  Interacting cases will simply lack some of the higher-dimension conditions (e.g., $p^2=0$).
	
	For example, we can translate the most well-known semi-shortening conditions $(\hat{Q}^i)^2 \left| 0 \right>_{\dot{\alpha}_1\cdots\dot{\alpha}_{2\bar{j}}} = 0$ and $\epsilon^{\alpha\beta}\hat{Q}^i_{~\beta}\left| 0 \right>_{\alpha\alpha_2\cdots\alpha_{2j},\dot{\alpha}_1\cdots\dot{\alpha}_{2\bar{j}}} = 0$  into superspace language as $(q^i)^2\phi_{\dot{\alpha}_1\cdots\dot{\alpha}_{2\bar{j}}} = 0$ and $\epsilon^{\alpha\beta}q^i_{~\beta}\phi_{\alpha\alpha_2\cdots\alpha_{2j},\dot{\alpha}_1\cdots\dot{\alpha}_{2\bar{j}}} = 0$ respectively. In the paper by F. Dolan and H. Osborn\cite{Dolan:2002zh}, there is another semi-shortening  condition $(Q^i_2 - \frac{1}{2 j + 1} Q^i_1) \left| j , \bar{j} \right> = 0$ which is, in fact, just another form of $\epsilon^{\alpha\beta}\hat{Q}^i_{~\beta}\left| 0 \right>_{\alpha\alpha_2\cdots\alpha_{2j},\dot{\alpha}_1\cdots\dot{\alpha}_{2\bar{j}}} = 0$. In terms of superfields, this condition is equivalent to
	\begin{equation}
		\left(q^{i\alpha}m_{\alpha}^{~+} + j q^{i+}\right) \phi_{\alpha_1\cdots\alpha_{2j},\dot{\alpha}_1\cdots\dot{\alpha}_{2\bar{j}}} = 0.\label{special}
	\end{equation}
	Equation (\ref{special}) is a special case of constraint $\lbrace 13 \rbrace$, which is 
$$ q^{k\gamma}\left[  \delta^i_{~j}\left(m_{{\gamma}}^{~\alpha} - \frac{i}{2}\delta_{{\gamma}}^{~\alpha}(d-2i)\right) - \delta^{\alpha}_{~\gamma} r^i_{~j} \right] + (k \leftrightarrow i)=0 $$
by taking $k = i$ and $\alpha = +$. These conditions all come from constraint $\lbrace 6 \rbrace$, $q^{j\alpha}q^{j}_{~\alpha}=0$. We can also obtain the complex conjugate semi-shortening conditions by using constraint $\lbrace 3 \rbrace$.
	
	To conclude this section, we claim that the full set of possible semi-shortening conditions quadratic in covariant derivatives can be obtained by just analyzing field equations without using the unitarity condition.
	
	\section{Semi-shortening conditions\label{semishort}}
	
	We now generalize the method used in section \ref{onshell}.
	First we note that the full set of constraints quadratic in covariant derivatives can be expressed in manifestly covariant form as the equation\cite{Siegel:1987ik}
	\begin{equation}
		g_{(\mathcal{M}}^{~~(\mathcal{N}}g_\mathcal{P]}^{~~\mathcal{Q}]} = 0, \label{G2}
	\end{equation}
	where $(~]$ means it is antisymmetric when interchanging two fermionic indices and symmetric otherwise.  We define the set $g^2$ as the collection of all quadratic generators of this form. This set includes the massless Klein-Gordan equation $p^2 = 0$ in 4D spacetime. Thus, the results of that section could be obtained by looking for the covariant expression containing $p^2=0$.  This covariance is under transformations generated by covariant derivatives. (As for all covariant derivatives, these equations are {\it invariant} under superconformal symmetry transformations.)  Thus, taking the (anti)commutator of almost any one of $g^2$ with $g_{\alpha}^{~i}$ or $g_i^{~\dot{\alpha}}$ gives other constraints in this set.  In general,
	\begin{eqnarray}
		\left[ g_\mathcal{M}^{~~\mathcal{N}}, g_{(\mathcal{P}}^{~~(\mathcal{Q}}g_{\mathcal{R}]}^{~~\mathcal{S}]}\right\rbrace & = & \delta^{\mathcal{N}}_{\mathcal{P}}g_{(\mathcal{M}}^{~~(\mathcal{Q}}g_{\mathcal{R}]}^{~~\mathcal{S}]} + (-1)^{\kappa \left((\mathcal{M}+\mathcal{N})(\mathcal{P}+\mathcal{Q})\right)}\delta^{\mathcal{N}}_{\mathcal{R}}g_{(\mathcal{P}}^{~~(\mathcal{Q}}g_{\mathcal{M}]}^{~~\mathcal{S}]}\nonumber\\
		& & - (-1)^{\kappa \left((\mathcal{M}+\mathcal{N})(\mathcal{P}+\mathcal{Q})\right)}\delta^{\mathcal{Q}}_{\mathcal{M}}g_{(\mathcal{P}}^{~~(\mathcal{N}}g_{\mathcal{R}]}^{~~\mathcal{S}]} \nonumber\\
		& & - (-1)^{\kappa \left((\mathcal{M}+\mathcal{N})(\mathcal{P}+\mathcal{Q}+\mathcal{R}+\mathcal{S})\right)} \delta^{\mathcal{S}}_{\mathcal{M}}g_{(\mathcal{P}}^{~~(\mathcal{Q}}g_{\mathcal{R}]}^{~~\mathcal{N}]}.\label{G2CR}
	\end{eqnarray}
	For example, if we start with $g_{(i}^{~[\alpha}g_{i)}^{~\beta]}=0$ (i.e., $(q^i)^2 = 0$) together with superconformal generators leads to $g_{(i}^{~(\gamma}g_{i)}^{~j)}=0$, $g_{[i}^{~(\alpha}g_{\gamma]}^{~\beta]}=0$, $g_{(i}^{~(k}g_{\gamma]}^{~\rho]}=0$, and $g_{(i}^{~(k}g_{i)}^{~j)}=0$.
	
	Of course, all the shortening conditions form a subset of the set of all generators $g^1$. Since the generators and the indices will increase rapidly as we go on and it is not important here to know what the indices and the coefficients are, we will only give qualitative expressions of the (anti)commutation relations unless otherwise needed. For example, we will write equation (\ref{G2CR}) as
	\begin{equation*}
		\left[ g , g^2 \right\rbrace \sim \delta g^2.
	\end{equation*}
	
	The next thing to check is the (anti)commutation relation of any two elements in $g^2$. It can be easily found by using the following identity:
	\begin{equation}
		\left[ g_{(\mathcal{M}}^{~~(\mathcal{N}}g_{\mathcal{P}]}^{~~\mathcal{Q}]}, \mathcal{O} \right\rbrace = 2 g_{(\mathcal{M}}^{~~(\mathcal{N}}\left[g_{\mathcal{P}]}^{~~\mathcal{Q}]}, \mathcal{O}\right\rbrace - \left[ g_{(\mathcal{M}}^{~~(\mathcal{N}},\left[g_{\mathcal{P}]}^{~~\mathcal{Q}]}, \mathcal{O}\right\rbrace\right\rbrace,\label{identity}
	\end{equation}
	where $\mathcal{O}$ is an arbitrary operator. Therefore, by substituting $\mathcal{O}$ with some element in $g^2$ together with equation (\ref{G2}) we get
	\begin{equation}
		\left[g^2,g^2\right\rbrace \sim \delta g \left(g^2\right) + \delta \delta g^2. \label{G2G2}
	\end{equation}
	The $g \left(g^2\right)$ term means a symmetry generator ``times'' an element in $g^2$ that cannot be combined into $g^3$, the set of all cubic operator of the form $g_{(\mathcal{M}}^{~~(\mathcal{N}}g_{\mathcal{P}}^{~~\mathcal{Q}}g_{\mathcal{R}]}^{~~\mathcal{S}]}$. Equation (\ref{G2G2}) tells us that a superfield under some constraints in $g^2$ can only give constraints the same strength as or weaker than $g^2$, it never goes to $g^1$. In other words, no matter how many semi-shortening conditions there are, it won't imply any shortening conditions.
	
	From the discussion above, we found that $g^1$ and $g^2$ have some nice features: They are closed under symmetry transformation and they don't give stronger constraints ($g^1$ is the strongest set of constraints other than making the field identically zero). The question now arises: Does $g^3$ have these properties?  Before checking $\left[ g^3, g^3\right\rbrace$, we first derive an ``intermediate step'', $\left[ g^2, g^3\right\rbrace$, which is of the same importance as $\left[ g^3, g^3\right\rbrace$. By using equation (\ref{identity}), we have the following:
	\begin{equation}
		\left[ g^2, g^3\right\rbrace \sim \delta g (g^3) + \delta \delta (g^3). \label{G2G3}
	\end{equation}
	Since we are interested in $\left[ g^3, g^3\right\rbrace$ at the first place, we will come back to the equation (\ref{G2G3}) later. With the aid of equation (\ref{G2G3}), we get the following:
	\begin{equation}
		\left[ g^3, g^3\right\rbrace \sim \delta (g^2) (g^3) + \delta \delta g (g^3) + \delta\delta\delta(g^3). \label{G3G3}
	\end{equation}
	From the equation above, we can conclude that $\left[ g^3, g^3\right\rbrace$ won't imply any constraint stronger than $g^3$.
	
	Back to the ``intermediate step'', equation (\ref{G2G3}). One may notice that the (anti)commutation relation between $g^2$ and $g^3$ gives constraints same as or weaker than $g^3$ (also $g^1$ with $g^3$ gives $g^3$). This means weak constraints always stay weak or even weaker, and it will not effect stronger constraints.
	
	The above statements can be generalized to all $g^n$ with positive and finite integer $n$, where 
	$$ g^n \equiv \left\lbrace g_{(\mathcal{ M }_1}^{~~(\mathcal{ N }_1}  g_{\mathcal{ M }_2}^{~~\mathcal{ N }_2} \cdots g_{\mathcal{ M }_n]}^{~~\mathcal{ N }_n]}\right\rbrace $$
	The first thing to do is to find the (anti)commutation relations between elements in two arbitrary sets, $g^n$ and $g^m$. To find the (anti)commutation relation between the elements of these sets, we first generalize equation (\ref{identity}) to the $n^{\text{th}}$ power:
	\begin{eqnarray}
		\left[ g^n , \mathcal{O}\right\rbrace & = & \begin{pmatrix}n\\1\end{pmatrix} g_{(\mathcal{A}_1}^{~~(\mathcal{B}_1} g_{\mathcal{A}_2}^{~~\mathcal{B}_2} \cdots \left[ g_{\mathcal{A}_n]}^{~~\mathcal{B}_n]}, \mathcal{O}\right\rbrace\nonumber\\ 
		& & + \left(-1 \right)^{1} \begin{pmatrix}n\\2\end{pmatrix} g_{(\mathcal{A}_1}^{~~(\mathcal{B}_1} g_{\mathcal{A}_2}^{~~\mathcal{B}_2} \cdots \left[ g_{\mathcal{A}_{n-1}]}^{~~\mathcal{B}_{n-1}]}, \left[ g_{\mathcal{A}_n]}^{~~\mathcal{B}_n]}, \mathcal{O}\right\rbrace\right\rbrace + \cdots\nonumber\\
		& & + \left(-1 \right)^{n-1} \begin{pmatrix}n\\n\end{pmatrix} \left[g_{(\mathcal{A}_1}^{~~(\mathcal{B}_1}, \left[ g_{\mathcal{A}_2}^{~~\mathcal{B}_2}, \cdots \left[ g_{\mathcal{A}_{n-1}]}^{~~\mathcal{B}_{n-1}]}, \left[ g_{\mathcal{A}_n]}^{~~\mathcal{B}_n]}, \mathcal{O}\right\rbrace\right\rbrace\right\rbrace\right\rbrace \nonumber\\
		& = & \sum_{i=1}^n \left(-1\right)^{i-1}\begin{pmatrix}n\\i\end{pmatrix}\tilde{g}^{n-i}{\text{ad} _{\tilde{g}}}^{i}\mathcal{O},\label{identityn}
	\end{eqnarray}
	where 
$$ \tilde{g}^{n-i}{\text{ad} _{\tilde{g}}}^{i} = g_{(\mathcal{A}_1}^{~~(\mathcal{B}_1}  \cdots g_{\mathcal{A}_{n-i}}^{~~\mathcal{B}_{n-i}}\left[g_{\mathcal{A}_{n-i+1}}^{~~\mathcal{B}_{n-i+1}}, \cdots \left[ g_{\mathcal{A}_{n-1}}^{~~\mathcal{B}_{n-1}}, \left[ g_{\mathcal{A}_n]}^{~~\mathcal{B}_n]}, \mathcal{O}\right\rbrace\right\rbrace\right\rbrace. $$
The proof is in appendix \ref{mathinduction}.
	
	Without loss of generality, we assume $m\geq n$ and substitute $\mathcal{O}$ with $g^m$. By using equation (\ref{identityn}), the (anti)commutation relation between $g^n$ and $g^m$ is
	\begin{equation}
		\left[ g^n , g^m \right\rbrace \sim \delta g^{n-1} g^m + \delta\delta g^{n-2}g^m + \cdots + \underbrace{\delta \delta \cdots \delta}_n g^m.\label{algebra}
	\end{equation}
	From this relation, we conclude the stronger constraints transform weaker constraints into some other weaker constraints but not the other way around.
	
	\section{Comparison with the ``old'' results\label{comparison}}
	
	In this section, we show that the semi-shortening conditions in F. Dolan and H. Osborn's paper\cite{Dolan:2002zh} can be reproduced by using $g^2$ and $g^3$ constraints. As has been discussed in section \ref{onshell}, 
	$$ (q_i)^2 \phi_{\dot{\alpha_1}\cdots\dot{\alpha}_{2\bar{j}}} = 0 \hbox{\quad and\quad} 
	q_{i}^{~\alpha}\phi_{\alpha\alpha_2\cdots\alpha_{2j},\dot{\alpha_1}\cdots\dot{\alpha}_{2\bar{j}}} = 0 $$ 
(and the complex conjugate of that) are just special cases of $g^2$ constraints. The rest of the semi-shortening conditions in the paper are 
	\begin{eqnarray*}
		\left\lbrace
		\begin{array}{ll}
			p^{\dot{\alpha}\alpha}\phi_{\alpha\alpha_2\cdots\alpha_{2j},\dot{\alpha}\dot{\alpha}_2\cdots\dot{\alpha}_{2\bar{j}}}=0 & \hbox{\quad (with scale dimension\quad} \Delta = 2 + j + \bar{j})\cr
			p^{\dot{\alpha}\alpha}q_{i\alpha}\phi_{\dot{\alpha}\dot{\alpha}_2\cdots\dot{\alpha}_{2\bar{j}}}=0 & \hbox{\quad (with\quad} \Delta = 2 + \bar{j})\cr
			p^{\dot{\alpha}\alpha}\bar{q}_{\dot{\alpha}}^{~i}\phi_{\alpha\alpha_2\cdots\alpha_{2j}}=0 & \hbox{\quad (with\quad} \Delta = 2 + j\cr
			p^{\dot{\alpha}\alpha}\left[q_{i\alpha} , \bar{q}_{\dot{\alpha}}^{~j}\right]\phi=0 & \hbox{\quad (with\quad} r = 0)
		\end{array}\right.
	\end{eqnarray*}
	in superspace language. These are actually special cases of $g^3$-constraints acting on different superfields.
	
	Take $p^{\dot{\alpha}\alpha}\left[q_{i\alpha} , \bar{q}_{\dot{\alpha}}^{~j}\right]\phi=0$ as example. It can be written as $g_{(\dot{\alpha}}^{~(\alpha}g_{\dot{\beta}}^{~\beta}g_{i]}^{~j]}\phi = 0$ if $\phi$ satisfies $r=0$. The detailed derivation is shown in the following:
	\begin{eqnarray*}
		0 & = & g_{(\dot{\alpha}}^{~(\alpha}g_{\dot{\beta}}^{~\beta}g_{i]}^{~j]}\phi\\
		& = & - 3 \left( p_{[\dot{\alpha}}^{~[\alpha}q_{|i|}^{~\beta]}\bar{q}_{\dot{\beta}]}^{~j} - p_{[\dot{\alpha}}^{~[\alpha}\bar{q}_{\dot{\beta}]}^{~|j|}q_{i}^{~\beta]} - p_{[\dot{\alpha}}^{~[\alpha}p_{\dot{\beta}]}^{~\beta]}r_{i}^{~j}\right)\phi\\
		& = & - 3 \left( p_{[\dot{\alpha}}^{~[\alpha}q_{|i|}^{~\beta]}\bar{q}_{\dot{\beta}]}^{~j} - p_{[\dot{\alpha}}^{~[\alpha}\bar{q}_{\dot{\beta}]}^{~|j|}q_{i}^{~\beta]}\right)\phi\\
		\Rightarrow 0 & = & p^{\dot{\alpha}\alpha}\left[q^i_{~\alpha} , \bar{q}_{j\dot{\alpha}}\right]\phi.
	\end{eqnarray*}The relations between $g^3$ constraints and the semi-shortening conditions in their paper are listed in the following table:
\vskip.2in
	\begin{tabular}{|l|l|l| }
		\hline
		$g^3$ & Dolan and Osborn & Shortening conditions \\
		\hline
		$g_{(\dot{\alpha}}^{~(\alpha}g_{\dot{\beta}}^{\beta}g_{i]}^{~j]}$ & $p^{\dot{\alpha}\alpha}\left[q_{i\alpha} , \bar{q}_{\dot{\alpha}}^{~j}\right]\phi=0$ & R-symmetry eigenvalue $=0$\\
		\hline
		$g_{(\dot{\alpha}}^{~(\alpha}g_{\dot{\beta}}^{~\beta}g_{i]}^{~\dot{\gamma}]}$ & $p^{\dot{\alpha}\alpha}q_{i\alpha}\phi_{\dot{\alpha}\dot{\alpha}_2\cdots\dot{\alpha}_{2\bar{j}}}=0$ & $\Delta = 2 + \bar{j}$, $\dot{\gamma} = -$\\
		\hline
		$g_{(\alpha}^{~(\beta}g_{\dot{\alpha}}^{\gamma}g_{\dot{\beta}]}^{~i]}$ & $p^{\dot{\alpha}\alpha}\bar{q}_{\dot{\alpha}}^{~i}\phi_{\alpha\alpha_2\cdots\alpha_{2j}}=0$ & $\Delta = 2 + j$, $\gamma = +$\\
		\hline
		$g_{(\alpha}^{~(\beta}g_{\dot{\alpha}}^{\gamma}g_{\dot{\beta}]}^{~\dot{\gamma}]}$ & $p^{\dot{\alpha}\alpha}\phi_{\alpha\alpha_2\cdots\alpha_{2j},\dot{\alpha}\dot{\alpha}_2\cdots\dot{\alpha}_{2\bar{j}}}=0$ & $\Delta = 2 + j + \bar{j}$, $\beta = +$, $\dot{\gamma} = -$\\
		\hline
	\end{tabular}
\vskip.2in	
	In fact, we can get the whole list of constraints by starting with the first constraint ($g_{(\dot{\alpha}}^{~(\alpha}g_{\dot{\beta}}^{\beta}g_{i]}^{~j]} = 0$) and repetitively taking (anti)commutators with $s$ or $\bar{s}$. We can get the second constraint ($g_{(\dot{\alpha}}^{~(\alpha}g_{\dot{\beta}}^{~\beta}g_{i]}^{~\dot{\gamma}]}=0$) or the third constraint ($g_{(\alpha}^{~(\beta}g_{\dot{\alpha}}^{\gamma}g_{\dot{\beta}]}^{~i]}$ = 0) by applying an $s$ or $\bar{s}$ on the first constraint. By applying both $s$ and $\bar{s}$ once, one can get $g_{(\alpha}^{~(\beta}g_{\dot{\alpha}}^{\gamma}g_{\dot{\beta}]}^{~\dot{\gamma}]}=0$. The full constraints induced by $g_{(\dot{\alpha}}^{~(\alpha}g_{\dot{\beta}}^{\beta}g_{i]}^{~j]} = 0$ are listed in appendix \ref{g3constraints}.
	
	Here we should also mention that the constraints induced by $g_{(\dot{\alpha}}^{~(\alpha}g_{\dot{\beta}}^{\beta}g_{i]}^{~j]} = 0$ form a closed set. One might expect that some other constraints will be induced by the (anti)commutation relation between two arbitrary $g^3$-constraints. However, according to equation (\ref{algebra}), the (anti)commutation relation between $g^3$-constraints will be ``proportional'' to $g^3$. In other words, since $g_{(\dot{\alpha}}^{~(\alpha}g_{\dot{\beta}}^{\beta}g_{i]}^{~j]} = 0$ already induced all possible $g^3$-constraints, the (anti-) commutation relation is ``proportional'' to some $g^3$-constraint. Hence, it will not give additional constraints.
	
	As advertised, we have reproduced all the semi-shortening constraints by using $g^2$ and $g^3$ constraints. To this day, only $g^2$ and $g^3$ constraints have been considered in the literature. Our work shows that there can be infinite numbers of semi-shortening constraints (i.e. $g^n$'s) which we think are complete, in the sense that any set of semi-shortening conditions must be a subset of them. The following section is an explicit example of $g^{n+1}$ constraints satisfied by tr$\varphi^n$. We expressed all the constraints on a multiplet, including those on the Lorentz and $SU(4)$ representations, as differential equations on coset space.

	\section{$\mathcal{N}=4$ SYM in projective superspace\label{N=4}}
	
	The generalized semi-shortening conditions ($g^n=0$) can be used on the $\mathcal{N}=4$ SYM field strength in projective superspace. In general, the field strength $\varphi$ obeys semi-shortening conditions 
	\begin{align}
		r_{(a}^{~(b}r_{c)}^{~d)}\varphi = 0\quad (g_{(a}^{~(b}g_{c)}^{~d)}\varphi = 0).\label{rr}
	\end{align}
In the free theory, this generalizes to all the $g^2$ constraints, but for the nonabelian case the derivatives must be generalized to gauge-covariant derivatives, and ``nonminimal" field strength terms are needed.  However, no nonminimal terms are needed for the above equation, since the $r$ derivatives have dimension 0, whereas field strengths have dimension of at least 1.  (Furthermore, a gauge can be chosen where the gauge potential for $r$ vanishes.)

 A direct consequence of this for the BPS operators is that 
 $$ r^{n+1}\text{tr}\ \varphi^n = r_{(i_1}^{~(j_1} \cdots r_{i_{n+1})}^{~j_{n+1})}\text{tr}\ \varphi^n= 0 $$ 
 since at least one of the $\varphi$'s will be hit by two $r$'s. Also, note the $r$ derivatives always reduce to ordinary derivatives outside the trace, since it's a gauge singlet.  Since we are working with projective superspace, we divide R-symmetry indices into two categories ($i'$, $i''$) where the primed ones are antichiral and the double primed ones are chiral. The field strength $\varphi$ vanishes when hit with $q_{i'}^{~\alpha}$ and $\bar{q}_{\dot{\alpha}}^{~i''}$ ($g_{i'}^{~\alpha}$ and $g_{\dot{\alpha}}^{~i''}$). However, the semi-shortening condition above is not invariant under some supersymmetry transformations. Therefore, we can apply the algorithm discussed in section \ref{semishort} to find other semi-shortening conditions.
	
	Take $n = 3$ as an example,
	\begin{eqnarray}
		0 & = & g_{j'}^{~\alpha} g_{(a}^{~(b} g_{c}^{~d} g_{e}^{~f} g_{h)}^{~i)}\text{tr}\ \varphi^3\nonumber\\
		& = & \left[ g_{j'}^{~\alpha}, g_{(a}^{~(b} g_{c}^{~d} g_{e}^{~f} g_{h)}^{~i)}\right]\text{tr}\ \varphi^3 + g_{(a}^{~(b} g_{c}^{~d} g_{e}^{~f} g_{h)}^{~i)} g_{j'}^{~\alpha}\text{tr}\ \varphi^3\nonumber\\
		& = & \left(\delta_{j'}^{~b} g_{(a}^{~(\alpha} g_{c}^{~d} g_{e}^{~f} g_{h)}^{~i)} + \delta_{j'}^{~d} g_{(a}^{~(b} g_{c}^{~\alpha} g_{e}^{~f} g_{h)}^{~i)} + \delta_{j'}^{~f} g_{(a}^{~(b} g_{c}^{~d} g_{e}^{~\alpha} g_{h)}^{~i)}\right.\label{ex}\\
		& & \left.+ \delta_{j'}^{~i} g_{(a}^{~(b} g_{c}^{~d} g_{e}^{~f} g_{h)}^{~\alpha)}\right)\text{tr}\ \varphi^3,\nonumber
	\end{eqnarray}
	where the unprimed Latin indices are arbitrary numbers from 1 to 4. It is obvious from equation (\ref{ex}) that $g_{(a}^{~(\alpha} g_{c}^{~d} g_{e}^{~f} g_{h)}^{~i)} \text{tr}\ \varphi^3 = 0$. Repeatedly applying $\left[ g_{i'}^{~\alpha}, ~ \cdot ~\right\rbrace$, $\left[ g_{\dot{\alpha}}^{~i''}, ~ \cdot ~\right\rbrace$, $\left[ g_{\alpha}^{~i}, ~ \cdot ~\right\rbrace$, or $\left[ g_{i}^{~\dot{\alpha}}, ~ \cdot ~\right\rbrace$ to all the constraints, we get the set of constraints induced by $g_{(a}^{~(b}g_{c}^{~d}g_{e}^{~f}g_{h)}^{~i)} = 0$, which is made of and only of all the positive scale dimension $g^4$ constraints.

	One might expect that there are some weaker constraints implied by taking the (anti)commutator of two arbitrary  constraints above. However, these weaker constraints can also be decomposed into three generators times some positive scale dimension constraints, therefore no additional constraints. For example, one of the constraints induced by 
	$$ g_{(a}^{~(b}g_{c}^{~d}g_{e}^{~f}g_{h)}^{~i)}=0 \hbox{\quad and\quad} g_{(j}^{~(k}g_{l}^{~m}g_{n}^{~o}g_{p)}^{~\alpha)}=0 $$
is
	\begin{eqnarray*}
		0 & = & g_{(a}^{~(\alpha}g_{b}^{~c}g_{d)}^{~e)}g_{(f}^{~(h}g_{i}^{~j}g_{k}^{~l}g_{m)}^{~n)}\text{tr}(\varphi^3) = g_{(a}^{~(\alpha}g_{b}^{~c}g_{d)}^{~e)}\left(g_{(f}^{~(h}g_{i}^{~j}g_{k}^{~l}g_{m)}^{~n)}\text{tr}(\varphi^3)\right)
	\end{eqnarray*}
	which gives nothing but $0=0$. Therefore, the shortening and semi-shortening constraints in this case, $g^4$,  form a closed set.
	
	{\it A general rule for projective superspace: If there exists a particular constraint $g^m \phi = 0$, this would imply all the positive scale dimension elements in $g^m$ to be constraints on $\phi$; unless this $g^m$ has at least one R-symmetry index that is not arbitrary.} 
	
	The $n=3$ discussion above is an example of this rule.
	
	Since the constraint $r^{n+1}\text{tr}\ \varphi^n = r_{(i_1}^{~(j_1} \cdots r_{i_{n+1})}^{~j_{n+1})}\text{tr}\ \varphi^n= 0$ is always true for arbitrary $i$'s and $j$'s, by using the above mentioned rule, 
	\begin{align}
		\hbox{\it all\ } g^{n+1}\text{tr}\ \varphi^n = 0. \label{newresult}
	\end{align}
	
	Therefore, we got constraints for $\text{tr}\ \varphi^n$ by using semi-shortening constraints satisfied by $\varphi$. One can get the explicit form of the constraint by simply expand it and rewrite everything in covariant derivatives.
	
	We take $n=3$ in equation (\ref{newresult}) as an example. We can choose $g^4$ to be $g_{(\dot{\rho}}^{~(\alpha}g_{\dot{\sigma}}^{~\beta}g_{i}^{~j}g_{k]}^{~l]}$, which can be expanded as follows (together with equation (\ref{CR})):
	\begin{align*}
		0 = & g_{(\dot{\rho}}^{~(\alpha}g_{\dot{\sigma}}^{~\beta}g_{i}^{~j}g_{k]}^{~l]}\text{tr} \varphi^3\\
		= & \Big[g_{\dot{\rho}}^{~\alpha} g_{\dot{\sigma}}^{~\beta}\left(6g_{i}^{~j}g_{k}^{~l} + \delta_{i}^{~j}\delta_{k}^{~l}\right) + 12 g_{\dot{\rho}}^{~\alpha} \left[g_{\dot{\sigma}}^{~j},g_{i}^{~\beta}\right]g_{k}^{~l} - 3 \left\lbrace g_{\dot{\rho}}^{~j}g_{\dot{\sigma}}^{~l},g_{i}^{~\alpha}g_{k}^{~\beta} \right\rbrace\\
		& + (i\leftrightarrow k) + (j\leftrightarrow l) - (\dot{\rho}\leftrightarrow \dot{\sigma})- (\alpha\leftrightarrow \beta)\Big] \text{tr} \varphi^3.
	\end{align*}
	Rewrite $g$ in terms of individual covariant derivatives (see equation (\ref{matrix})):
	\begin{align*}
		0 = & \Big[p_{\dot{\rho}}^{~\alpha} p_{\dot{\sigma}}^{~\beta}\left(6r_{i}^{~j}r_{k}^{~l} + \delta_{i}^{~j}\delta_{k}^{~l}\right) + 12 p_{\dot{\rho}}^{~\alpha} \left[\bar{q}_{\dot{\sigma}}^{~j}, q_{i}^{~\beta}\right]r_{k}^{~l} - 3 \left\lbrace \bar{q}_{\dot{\rho}}^{~j}\bar{q}_{\dot{\sigma}}^{~l},q_{i}^{~\alpha}q_{k}^{~\beta} \right\rbrace\\
		& + (i\leftrightarrow k) + (j\leftrightarrow l) - (\dot{\rho}\leftrightarrow \dot{\sigma})- (\alpha\leftrightarrow \beta)\Big] \text{tr}\ \varphi^3\\
		= & - \epsilon^{\alpha\beta}\epsilon_{\dot{\rho}\dot{\sigma}}\bigg[p^{\dot{\gamma}\delta} p_{\dot{\gamma}\delta}\left(6r_{(i}^{~(j}r_{k)}^{~l)} + \delta_{(i}^{~(j}\delta_{k)}^{~l)}\right)+ 12 p^{\dot{\gamma}\delta} \left[\bar{q}_{\dot{\gamma}}^{~(j}, q_{(i|\delta}\right]r_{|k)}^{~l)}\\
		& - 3 \left\lbrace \bar{q}^{\dot{\gamma}(j}\bar{q}_{\dot{\gamma}}^{~l)},q_{(i}^{~\delta}q_{k)\delta} \right\rbrace\bigg] \text{tr} \varphi^3.
	\end{align*}
	Therefore, we found tr$\varphi^3$ satisfies semi-shortening constraint:
	\begin{align*}
		0 = &\bigg[p^{\dot{\gamma}\delta} p_{\dot{\gamma}\delta}\left(6r_{(i}^{~(j}r_{k)}^{~l)} + \delta_{(i}^{~(j}\delta_{k)}^{~l)}\right)+ 12 p^{\dot{\gamma}\delta} \left[\bar{q}_{\dot{\gamma}}^{~(j}, q_{(i|\delta}\right]r_{|k)}^{~l)}\\
		& - 3 \left\lbrace \bar{q}^{\dot{\gamma}(j}\bar{q}_{\dot{\gamma}}^{~l)},q_{(i}^{~\delta}q_{k)\delta} \right\rbrace\bigg] \text{tr} \varphi^3.
	\end{align*}
	
	We can also choose $g^5$ semi-shortening constraints on tr$\varphi^4$. Here we choose $g^5$ as follows:
	\begin{eqnarray*}
		g_{(\dot{\rho}}^{~(\alpha}g_{\dot{\sigma}}^{~\beta}g_{i}^{~j}g_{k}^{~l}g_{m]}^{~n]}  & = & 10g_{[\dot{\rho}}^{~[\alpha}g_{\dot{\sigma}]}^{~\beta]}g_{i}^{~j}g_{k}^{~l}g_{m}^{~n} - 60g_{[\dot{\rho}|}^{~[\alpha}g_{i}^{~\beta]}g_{|\dot{\sigma}]}^{~j}g_{k}^{~l}g_{m}^{~n} - 30g_{i}^{~[\alpha}g_{k}^{~\beta]}g_{[\dot{\rho}}^{~j}g_{\dot{\sigma}]}^{~l}g_{m}^{~n}\\ 
 & &+ 30g_{[\dot{\rho}}^{~[\alpha}g_{\dot{\sigma}]}^{~\beta]}g_{i}^{~j}g_{k}^{~l}\delta_{m}^{~n} - 60g_{[\dot{\rho}|}^{~[\alpha}g_{i}^{~\beta]}g_{|\dot{\sigma}]}^{~j}g_{k}^{~l}\delta_{m}^{~n} + 35g_{[\dot{\rho}}^{~[\alpha}g_{\dot{\sigma}]}^{~\beta]}g_{i}^{~j}\delta_{k}^{~l}\delta_{m}^{~n}\\ 
 & &- 30g_{[\dot{\rho}|}^{~[\alpha}g_{i}^{~\beta]}g_{|\dot{\sigma}]}^{~j}\delta_{k}^{~l}\delta_{m}^{~n} + 15g_{[\dot{\rho}}^{~[\alpha}g_{\dot{\sigma}]}^{~\beta]}\delta_{i}^{~j}\delta_{k}^{~l}\delta_{m}^{~n} + (i \leftrightarrow k \leftrightarrow m)\\
 & & + (j \leftrightarrow l \leftrightarrow n),
	\end{eqnarray*}
	which indicates tr$\varphi^4$ satisfies the following constraint:
	\begin{eqnarray*}
		0 & = & \bigg[ p^{\dot{\alpha}\alpha} p_{\dot{\alpha}\alpha} \Big(2 r_{(i}^{~(j} r_k^{~l} r_{m)}^{~n)} + 6 r_{(i}^{~(j} r_k^{~l} \delta_{m)}^{~n)} + 7 r_{(i}^{~(j} \delta_k^{~l} r\delta_{m)}^{~n)} + 3 \delta_{(i}^{~(j} \delta_k^{~l} \delta_{m)}^{~n)}\Big)\\
		& & - 6 p^{\dot{\alpha}\alpha} q_{(i|\alpha}\bar{q}_{\dot{\alpha}}^{~(j} \Big( 2 r_{|k}^{~l}r_{m)}^{~n)} + 2 r_{|k}^{~l}\delta_{m)}^{~n)} + \delta_{|k}^{~l}\delta_{m)}^{~n)}\Big)- 6 q_{(i}^{~\alpha} q_{k|\alpha} \bar{q}^{\dot{\alpha}(j}\bar{q}_{\dot{\alpha}}^{~l} r_{|m)}^{~n)} \bigg] \text{tr}\varphi^4 .
	\end{eqnarray*}
	
	The two examples above are satisfied on BPS representations.

	\section{Conclusions}
	
	In section \ref{semishort}, we proved operators $g^n \equiv \left\lbrace g_{(\mathcal{ M }_1}^{~~(\mathcal{ N }_1}  g_{\mathcal{ M }_2}^{~~\mathcal{ N }_2} \cdots g_{\mathcal{ M }_n]}^{~~\mathcal{ N }_n]}\right\rbrace$ transform covariantly (up to an overall coefficient) under $(P)SU(2,2|\mathcal{N})$ symmetry. From the discussions in sections \ref{short} and \ref{comparison}, we found that the most well-known shortening and semi-shortening conditions form a subset of $g^1$, $g^2$, and $g^3$. Since the new method treat semi-shortening constraints as covariant operators $g^n$ (which are essentially derivatives), together with the algebras in section \ref{semishort}, it is easier to manipulate with and write down the explicit expressions of semi-shortening conditions. In particular, we found in subsection \ref{N=4} for the case of ${\cal N}$=4 SYM that the full set of $g^{n+1}$ constraints apply to the BPS operators tr\ $\varphi^n$ and gave some examples with explicit forms.

	\appendix
	\section{Appendix: Constraints from $p^2=0$\label{App}}
	
\noindent $\lbrace\ 0\ \rbrace\ \thinspace p^2 = 0$
\begin{enumerate}[$\lbrace$ 1 $\rbrace$]
	\item $p^{\dot{\alpha}\alpha}\bar{q}_{i\dot{\alpha}}=0$
	\item $p^{\dot{\alpha}\alpha}q^i_{~{\alpha}}=0$
	\item $\bar{q}_i^{~\dot{\alpha}}\bar{q}_{j\dot{\alpha}}=0$
	\item $q^{j\alpha}\bar{q}_i^{~\dot{\alpha}} + 2\delta^j_{~i}p^{\dot{\gamma}\alpha}m^{\dot{\alpha}}_{~\dot{\gamma}}
			+i\delta^j_{~i}p^{\dot{\alpha}\alpha}d-2p^{\dot{\alpha}\alpha}r^j_{~i}=0$
	\item $-\bar{q}_j^{~\dot{\alpha}}q^{i\alpha} + 2\delta^i_{~j}p^{\dot{\alpha}\gamma}m_{\gamma}^{~\alpha}-i\delta^i_{~j}p^{\dot{\alpha}\alpha}d-2p^{\dot{\alpha}\alpha}r^i_{~j}=0$
	\item $q^{i\alpha}q^{j}_{~\alpha}=0$
	\item $0=0$ (no new constraint)
	\item $\left\lbrace \bar{q}^{~\dot{\alpha}}_{i}\left[ \delta^k_{~j}\left(m^{\dot{\beta}}_{~\dot{\gamma}} + \frac{i}{2}\delta^{\dot{\beta}}_{~\dot{\gamma}}(d-2i)\right) - \delta^{\dot{\beta}}_{~\dot{\gamma}} r^k_{~j}\right]+ (i \leftrightarrow j ) \right\rbrace =0$
	\item $\left\lbrace \bar{q}^{~\dot{\alpha}}_{i}\epsilon^{\alpha\gamma}\left[ \delta^j_{~k}\left(m_{{\gamma}}^{~\beta} - \frac{i}{2}\delta_{{\gamma}}^{~\beta}d\right) - \delta^{\dot{\beta}}_{~\dot{\gamma}} r^j_{~k}\right]\right.$\\
			$\left. + \bar{q}^{~\dot{\gamma}}_{k}\epsilon^{\alpha\beta}\left[ \delta^j_{~i}\left(m^{\dot{\alpha}}_{~\dot{\gamma}} + \frac{i}{2}\delta^{\dot{\alpha}}_{~\dot{\gamma}}(d-2i)\right) - \delta^{\dot{\alpha}}_{~\dot{\gamma}} r^j_{~i}\right] \right\rbrace=0$
	\item $q^{j\alpha}\epsilon^{\dot{\alpha}\dot{\gamma}}\left[ \delta^k_{~i}\left(m^{\dot{\beta}}_{~\dot{\gamma}} + \frac{i}{2}\delta^{\dot{\beta}}_{~\dot{\gamma}}d\right) - \delta^{\dot{\beta}}_{~\dot{\gamma}} r^k_{~i}\right] - (j,\dot{\alpha} \leftrightarrow k,\dot{\beta} ) =0$
	\item $\left\lbrace \bar{q}^{~\dot{\alpha}}_{j}\epsilon^{\alpha\gamma}\left[ \delta^i_{~k}\left(m_{{\gamma}}^{~\beta} - \frac{i}{2}\delta_{{\gamma}}^{~\beta}d\right) - \delta^{\dot{\beta}}_{~\dot{\gamma}} r^i_{~k}\right]\right. - (j,\alpha \leftrightarrow k,\beta)=0$
	\item $q^{i\alpha}\epsilon^{\dot{\alpha}\dot{\gamma}}\left[ \delta^k_{~j}\left( m^{\dot{\beta}}_{~\dot{\gamma}} +\frac{i}{2}\delta^{\dot{\beta}}_{\dot{~\gamma}}d \right) -  \delta^{\dot{\beta}}_{\dot{~\gamma}}r^k_{~j} \right] $\\ $+ q^{k\gamma}\epsilon^{\dot{\alpha}\dot{\beta}}\left[  \delta^i_{~j}\left(m_{{\gamma}}^{~\alpha} - \frac{i}{2}\delta_{{\gamma}}^{~\alpha}(d-2i)\right) - \delta^{\alpha}_{~\gamma} r^i_{~j} \right]=0$
	\item $q^{k\gamma}\left[  \delta^i_{~j}\left(m_{{\gamma}}^{~\alpha} - \frac{i}{2}\delta_{{\gamma}}^{~\alpha}(d-2i)\right) - \delta^{\alpha}_{~\gamma} r^i_{~j} \right] + (k \leftrightarrow i)=0$
	\item $0=0$ (no new constraint)
	\item $0=0$ (no new constraint)
	\item $\epsilon^{\dot{\alpha}\dot{\gamma}}\left[ \delta^\ell_{~i}\left( m^{\dot{\rho}}_{~\dot{\alpha}} + \frac{i}{2}\delta^{\dot{\rho}}_{~\dot{\alpha}}d \right) - \delta^{\dot{\rho}}_{~\dot{\alpha}} r^\ell_{~i}\right]\left[ \delta^k_{~j}\left( m^{\dot{\beta}}_{~\dot{\gamma}} + \frac{i}{2}\delta^{\dot{\beta}}_{~\dot{\gamma}}(d-2i) \right)- \delta^{\dot{\beta}}_{~\dot{\gamma}} r^k_{~j}\right]$\\
			$ + (i \leftrightarrow j))=0$
	\item $0=0$ (no new constraint)
	\item $\left[ \delta^\ell_{~i}\left( m^{\dot{\rho}\dot{\alpha}} + \frac{i}{2}\epsilon^{\dot{\rho}\dot{\alpha}}d \right) - \epsilon^{\dot{\rho}\dot{\alpha}} r^\ell_{~i}\right]\left[ \delta^k_{~j}\left( m^{\alpha\beta} - \frac{i}{2}\epsilon^{\alpha\beta}d \right)- \epsilon^{\alpha\beta} r^kl_{~j}\right]$\\
		 $+\epsilon^{\alpha\beta}\left[ \delta^\ell_{~k}\left( m^{\dot{\rho}\dot{\gamma}} + \frac{i}{2}\epsilon^{\dot{\gamma}\dot{\rho}}d \right) - \epsilon^{\dot{\gamma}\dot{\rho}} r^\ell_{~k}\right]\left[\delta^j_{~i}\left( m^{\dot{\alpha}}_{~\dot{\gamma}} + \frac{i}{2}\delta^{\dot{\alpha}}_{~\dot{\gamma}}(d-2i) \right)-\delta^{\dot{\alpha}}_{~\dot{\gamma}}r^j_{~i}\right]$\\ $=0$
	 \item $0=0$ (no new constraint)
	 \item $\left[ \delta^\ell_{~i}\left( m^{\dot{\rho}\dot{\alpha}} + \frac{i}{2}\epsilon^{\dot{\rho}\dot{\alpha}}d \right) - \epsilon^{\dot{\rho}\dot{\alpha}} r^\ell_{~i}\right]\left[ \delta^k_{~j}\left( m^{\alpha\beta} - \frac{i}{2}\epsilon^{\alpha\beta}d \right)- \epsilon^{\alpha\beta} r^k_{~j}\right] $\\ $- (\ell\dot{\alpha} \leftrightarrow k,\dot{\rho})=0$
	 \item $\left[ \delta^\ell_{~i}\left( m^{\dot{\rho}\dot{\alpha}} + \frac{i}{2}\epsilon^{\dot{\rho}\dot{\alpha}}d \right) - \epsilon^{\dot{\rho}\dot{\alpha}} r^\ell_{~i}\right]\left[ \delta^k_{~j}\left( m^{\alpha\beta} - \frac{i}{2}\epsilon^{\alpha\beta}d \right) - \epsilon^{\alpha\beta} r^k_{~j}\right] $\\ $- (i,\alpha \leftrightarrow  j,\beta)=0$
	 \item $0=0$ (no new constraint)
	 \item $\left[ \delta^i_{~\ell}\left( m^{\alpha\rho} - \frac{i}{2}\epsilon^{\alpha\rho}d \right) - \epsilon^{\alpha\rho} r^i_{~\ell}\right]\left[ \delta^k_{~j}\left( m^{\dot{\beta}\dot{\alpha}} + \frac{i}{2}\epsilon^{\dot{\alpha}\dot{\beta}}d \right) - \epsilon^{\dot{\alpha}\dot{\beta}} r^k_{~j}\right] $\\
		 $+  \epsilon^{\dot{\alpha}\dot{\beta}}\left[ \delta^k_{~\ell}\left( m^{\gamma\rho} + \frac{i}{2}\epsilon^{\gamma\rho}d \right) - \epsilon^{\gamma\rho} r^k_{~\ell}\right]\left[ \delta^i_{~j}\left( m_{\gamma}^{~\alpha} - \frac{i}{2}\delta_{\gamma}^{~\alpha}(d-2i) \right)- \delta_{\gamma}^{~\alpha} r^i_{~j}\right]$\\ $=0$
	 \item $\left[ \delta^i_{~\ell}\left( m_{\alpha}^{~\rho} - \frac{i}{2}\delta_{\alpha}^{~\rho}d \right)- \delta_{\alpha}^{~\rho} r^i_{~\ell}\right]\left[ \delta^j_{~k}\left( m^{\alpha\beta} + \frac{i}{2}\epsilon^{\alpha\beta}(d-2i) \right)- \epsilon^{\alpha\beta} r^j_{~k}\right]$ $=0$
	 \item $0=0$ (no new constraint)
\end{enumerate}
		
	\section{Appendix: Closure of shortening \label{q}}
	
	In this appendix, we will use equation (\ref{CR}) to prove the closure of shortening conditions induced by setting $g_i^{~\alpha}=0$. We let $i$ and $\alpha$ be a fixed value, and the remaining indices are arbitrary. The nontrivial (nonvanishing) commutation relations are:
	
	\begin{enumerate}
		\item $0 = \lbrace g_i^{~\alpha} , g_{\beta}^{~j}\rbrace = \delta_\beta^\alpha g_i^{~j} + \delta_i^jg_\beta^{~\alpha}$. This implies $g_i^{~j}$ and $g_\beta^{~\alpha}$ should also vanish.
		\item Since $g_i^{~j}=0$, then we have $0 = [ g_i^{~j} , g_i^{~\alpha} ] = \delta_i^jg_i^{~\alpha}$. $g_i^{~\alpha} = 0$ is our starting point, therefore, this gives no new constraint.
		\item $0 = [ g_{\beta}^{~k} , g_i^{~j} ] = \delta_i^kg_{\beta}^{~j}$. This doesn't give a new condition.
		\item $0 = [ g_i^{~\alpha} , g_\beta^{~\alpha} ] = \delta_\beta^\alpha g_i^{~\alpha}$. This is again the starting point.
		\item $0 = [ g_\beta^{~\alpha} , g_{\gamma}^{~j} ] = \delta_\gamma^\alpha g_{\beta}^{~j}$. No new condition.
	\end{enumerate}
	Therefore, the superfield vanishes under $g_i^{~\alpha}$, $g_i^{~\dot{\beta}}$, $g_{\beta}^{~j}$, $g_{\beta}^{~\dot{\beta}}$, $g_i^{~j}$, and $g_\beta^{~\alpha}$. It won't imply the vanishing of any other $g_i^{~\beta}$ or $g_{\dot{\alpha}}^{~i}$. The algebra of shortening condition $g_{\dot{\alpha}}^{~i}=0$ is the complex conjugate of the above ones.

	The consequence of this can be easily realized diagrammatically. We first write down the generator matrix with superconformal and special conformal generators vanishing:
	\begin{eqnarray*}
		\bordermatrix{
			~  & \alpha & 1 & 2 & \cdots & \mathcal{N} & \dot{\alpha}\cr
			\beta & g_{\beta}^{~\alpha} & 0 & 0 & 0 & 0 & 0\cr
			1 & g_{1}^{~\alpha} & g_{1}^{~1} & g_{1}^{~2} & \cdots & g_{1}^{~\mathcal{N}} & 0 \cr
			2 & g_{2}^{~\alpha} & g_{2}^{~1} & g_{2}^{~2} & \cdots & g_{2}^{~\mathcal{N}} & 0 \cr
			\vdots & \vdots & & & \ddots & & \vdots \cr
			\mathcal{N} & g_{\mathcal{N}}^{~\alpha} & g_{\mathcal{N}}^{~1} & g_{\mathcal{N}}^{~2} & \cdots & g_{1}^{~\mathcal{N}} & 0 \cr
			\dot{\beta} & g_{\dot\beta}^{~\alpha} & g_{\dot\beta}^{~1} & g_{\dot\beta}^{~2} &\cdots & g_{\dot\beta}^{~\mathcal{N}} & g_{\dot\beta}^{~\dot{\alpha}}
		}.
	\end{eqnarray*}
	If we choose $g_i^{~\alpha} = 0$, then the whole row with such an element should completely vanish (also $g_\beta^{~\alpha}$):
		\begin{eqnarray*}
		\bordermatrix{
			~  & \alpha & 1 & 2 & \cdots & \mathcal{N} & \dot{\alpha}\cr
			\beta & \fbox{0} & 0 & 0 & 0 & 0 & 0\cr
			1 & g_{1}^{~\alpha} & g_{1}^{~1} & g_{1}^{~2} & \cdots & g_{1}^{~\mathcal{N}} & 0 \cr
			2 & g_{2}^{~\alpha} & g_{2}^{~1} & g_{2}^{~2} & \cdots & g_{2}^{~\mathcal{N}} & 0 \cr
			\vdots & \vdots & & & \ddots & & \vdots \cr
			i & \fbox{0} & \fbox{0} & \fbox{0} & \fbox{0} & \fbox{0} & 0 \cr
			\vdots & \vdots & & & \ddots & & \vdots \cr
			\mathcal{N} & g_{\mathcal{N}}^{~\alpha} & g_{\mathcal{N}}^{~1} & g_{\mathcal{N}}^{~2} & \cdots & g_{1}^{~\mathcal{N}} & 0 \cr
			\dot{\beta} & g_{\dot\beta}^{~\alpha} & g_{\dot\beta}^{~1} & g_{\dot\beta}^{~2} &\cdots & g_{\dot\beta}^{~\mathcal{N}} & g_{\dot\beta}^{~\dot{\alpha}}
		}.
	\end{eqnarray*}
	For $g_{\dot{\alpha}}^{~i} = 0$, instead of row, it is the column with $g_{\dot{\alpha}}^{~i}$ that vanishes (together with $g_{\dot{\alpha}}^{~\dot{\beta}}$).
	
	\section{Appendix: Proof of equation (\ref{identityn})\label{mathinduction}}
	
	In this section, we are going to prove the identity:
	\begin{eqnarray}
		\left[ g^n , \mathcal{O}\right\rbrace & = & \sum_{i=1}^n \left(-1\right)^{i-1}\begin{pmatrix}n\\i\end{pmatrix}\tilde{g}^{n-i}{\text{ad} _{\tilde{g}}}^{i}\mathcal{O},\label{identitym}
	\end{eqnarray}
where
	\begin{eqnarray*}
		\left\lbrace
		\begin{array}{l}
			\begin{pmatrix}m\\n\end{pmatrix} \equiv {\displaystyle\frac{(m+n)!}{m!n!}}\\
			\text{ad}_x y \equiv \left[ x, y\right]\\
			\tilde{g}^{n-i}{\text{ad} _{\tilde{g}}}^{i} = g_{(\mathcal{A}_1}^{~~(\mathcal{B}_1}  \cdots g_{\mathcal{A}_{n-i}}^{~~\mathcal{B}_{n-i}}\left[g_{\mathcal{A}_{n-i+1}}^{~~\mathcal{B}_{n-i+1}}\cdots \left[ g_{\mathcal{A}_{n-1}}^{~~\mathcal{B}_{n-1}}, \left[ g_{\mathcal{A}_n]}^{~~\mathcal{B}_n]} \mathcal{O}\right\rbrace\right\rbrace\right\rbrace
		\end{array}
		\right.
	\end{eqnarray*}
	
	This identity can be proven by using mathematical induction. Before starting this, it is useful to derive the equation:
	\begin{eqnarray}
		\text{ad} _{g^n} \mathcal{O} & = & \left[ g_{(\mathcal{A}_1}^{~~(\mathcal{B}_1}  \cdots g_{\mathcal{A}_{n}]}^{~~\mathcal{B}_{n}]} , \mathcal{O}\right\rbrace\nonumber\\
		& = & g_{(\mathcal{A}_1}^{~~(\mathcal{B}_1}\left[ g_{\mathcal{A}_{2}}^{~~\mathcal{B}_{2}} \cdots g_{\mathcal{A}_{n}]}^{~~\mathcal{B}_{n}]} ,\mathcal{O} \right\rbrace\nonumber\\
		& & + (-1)^{\kappa(\sum_{i=2}^n(\mathcal{A}_i+\mathcal{B}_i))\kappa(\mathcal{O})}\left[g_{(\mathcal{A}_1}^{~~(\mathcal{B}_1} ,\mathcal{O} \right\rbrace g_{\mathcal{A}_{2}}^{~~\mathcal{B}_{2}} \cdots g_{\mathcal{A}_{n}]}^{~~\mathcal{B}_{n}]}\nonumber\\
		& = & g_{(\mathcal{A}_1}^{~~(\mathcal{B}_1}\left[ g_{\mathcal{A}_{2}}^{~~\mathcal{B}_{2}} \cdots g_{\mathcal{A}_{n}]}^{~~\mathcal{B}_{n}]} ,\mathcal{O} \right\rbrace\nonumber\\
		& & + (-1)^{\kappa(\sum_{i=2}^n(\mathcal{A}_i+\mathcal{B}_i))\kappa(\mathcal{A}_1+\mathcal{B}_1)} g_{(\mathcal{A}_{2}}^{~~(\mathcal{B}_{2}} \cdots g_{\mathcal{A}_{n}}^{~~\mathcal{B}_{n}}\left[g_{(\mathcal{A}_1]}^{~~(\mathcal{B}_1]} ,\mathcal{O} \right\rbrace\nonumber\\
		& & + (-1)^{\kappa(\sum_{i=2}^n(\mathcal{A}_i+\mathcal{B}_i))\kappa(\mathcal{O})}\left[\left[g_{(\mathcal{A}_1}^{~~(\mathcal{B}_1} ,\mathcal{O} \right\rbrace, g_{\mathcal{A}_{2}}^{~~\mathcal{B}_{2}} \cdots g_{\mathcal{A}_{n}]}^{~~\mathcal{B}_{n}]}\right\rbrace\nonumber\\
		& = & g_{(\mathcal{A}_1}^{~~(\mathcal{B}_1}\left[ g_{\mathcal{A}_{2}}^{~~\mathcal{B}_{2}} \cdots g_{\mathcal{A}_{n}]}^{~~\mathcal{B}_{n}]} ,\mathcal{O} \right\rbrace + g_{(\mathcal{A}_{1}}^{~~(\mathcal{B}_{1}} \cdots g_{\mathcal{A}_{n-1}}^{~~\mathcal{B}_{n-1}}\left[g_{(\mathcal{A}_n]}^{~~(\mathcal{B}_n]} ,\mathcal{O} \right\rbrace\nonumber\\
		& & - \left[_{(\mathcal{A}_1}^{~~(\mathcal{B}_1}  \cdots g_{\mathcal{A}_{n-1}}^{~~\mathcal{B}_{n-1}},\left[g_{\mathcal{A}_{n}]}^{~~\mathcal{B}_{n}]} ,\mathcal{O} \right\rbrace\right\rbrace\nonumber\\
		& = & \tilde{g}~\text{ad} _{\tilde{g}^{n-1}}\mathcal{O} + \tilde{g}^{n-1}\text{ad} _{\tilde{g}}\mathcal{O} - \text{ad} _{\tilde{g}^{n-1}} \left(\text{ad} _{\tilde{g}} \mathcal{O}\right).\label{induction}
	\end{eqnarray}
	Now we start the proof:
	
	\begin{itemize}
	
	\item For $n = 2$, equation (\ref{identitym}) is obviously true since it is nothing but equation (\ref{identity}). (This can also be seen by taking $n=2$ in equation (\ref{induction}).)
	
	\item Assume equation (\ref{identitym}) is true for $n = k$. Then we can check if $n = k + 1$ is also true by direct calculation:
	
	\begin{eqnarray*}
		\text{ad} _{g^{k + 1}} \mathcal{O} & = & \tilde{g}^{k}\text{ad} _{\tilde{g}}\mathcal{O} + \tilde{g}~\text{ad} _{\tilde{g}^{k}}\mathcal{O} - \text{ad} _{\tilde{g}^{k}} \left(\text{ad} _{\tilde{g}} \mathcal{O}\right)\\
		& = & \tilde{g}^{k}\text{ad} _{\tilde{g}}\mathcal{O} + \tilde{g}\left[\sum_{i=1}^k \left(-1\right)^{i-1}\begin{pmatrix}k\\i\end{pmatrix}\tilde{g}^{k-i}{\text{ad} _{\tilde{g}}}^{i}\mathcal{O}\right]\\
		& & - \left[\sum_{i=1}^k \left(-1\right)^{i-1}\begin{pmatrix}k\\i\end{pmatrix}\tilde{g}^{k-i}{\text{ad} _{\tilde{g}}}^{i}\left(\text{ad} _{\tilde{g}}\mathcal{O}\right)\right]\\
		& = & \begin{pmatrix} k \\ 0\end{pmatrix} \tilde{g}^k \text{ad} _{\tilde{g}} \mathcal{O} + \begin{pmatrix} k \\ 1\end{pmatrix} \tilde{g}^k \text{ad} _{\tilde{g}} \mathcal{O} \\
		& & + \left[\sum_{i=2}^k \left(-1\right)^{i-1}\begin{pmatrix}k\\i\end{pmatrix}\tilde{g}^{k-i+1}{\text{ad} _{\tilde{g}}}^{i}\mathcal{O}\right]\\
		& & + \left[\sum_{i=2}^{k+1} \left(-1\right)^{i-1}\begin{pmatrix}k\\i-1\end{pmatrix}\tilde{g}^{k-i+1}{\text{ad} _{\tilde{g}}}^{i}\mathcal{O}\right]\\
		& = & \begin{pmatrix} k + 1 \\ 1\end{pmatrix} \tilde{g}^k \text{ad} _{\tilde{g}} \mathcal{O} + \left[\sum_{i=2}^{k+1} \left(-1\right)^{i-1}\begin{pmatrix}k + 1\\i\end{pmatrix}\tilde{g}^{k-i+1}{\text{ad} _{\tilde{g}}}^{i}\mathcal{O}\right]\\
		& = & \left[\sum_{i=1}^{k+1} \left(-1\right)^{i-1}\begin{pmatrix}k + 1\\i\end{pmatrix}\tilde{g}^{(k+1)-i}{\text{ad} _{\tilde{g}}}^{i}\mathcal{O}\right]
	\end{eqnarray*}
	where we have used equation (\ref{induction}) and $\begin{pmatrix} k + 1 \cr i \end{pmatrix} = \begin{pmatrix} k \cr i - 1\end{pmatrix} + \begin{pmatrix} k \cr i\end{pmatrix}$. Hence, equation (\ref{identitym}) is also true for $n = k+1$.
	
	\item By mathematical induction, equation (\ref{identitym}) is true for every integer $n\geq2$.
	\end{itemize}
	
	\section{Appendix: Full set of $g^3$-constraints.\label{g3constraints}}
	
	This appendix is the list of all possible $g^3$-constraints. This set can be induced by the highest scale dimension constraint: $g_{(\dot{\alpha}}^{~(\alpha}g_{\dot{\beta}}^{\beta}g_{i]}^{~j]} = 0$. Since negative scale dimension constraints always kill the superfield by construction ($s\phi = 0$, $\bar{s}\phi=0$, or $k\phi = 0$), we list only the constraints with non-negative scale dimension in the table below.
	
\vskip.2in	

	\begin{tabular}{|l | c | c | c | c | c |}
	\hline
	& $\alpha\beta i$ & $\alpha\beta \dot{\alpha}$ & $\alpha i j$ & $\alpha i\dot{\alpha}$ & $ijk$ \\
	\hline
	$\dot{\rho}\dot{\sigma}\ell$ & $g_{(\dot{\rho}}^{~(\alpha}g_{\dot{\sigma}}^{~\beta}g_{\ell]}^{~i]}$ & $g_{(\dot{\rho}}^{~(\alpha}g_{\dot{\sigma}}^{~\beta}g_{\ell]}^{~\dot{\alpha}]}$ & $g_{(\dot{\rho}}^{~(\alpha}g_{\dot{\sigma}}^{~i}g_{\ell]}^{~j]}$ & $g_{(\dot{\rho}}^{~(\alpha}g_{\dot{\sigma}}^{~i}g_{\ell]}^{~\dot{\alpha}]}$ & $g_{(\dot{\rho}}^{~(i}g_{\dot{\sigma}}^{~j}g_{\ell]}^{~k]}$\\
	\hline
	$\dot{\rho}\dot{\sigma}\rho$ & $g_{(\dot{\rho}}^{~(\alpha}g_{\dot{\sigma}}^{~\beta}g_{\rho]}^{~i]}$ & $g_{(\dot{\rho}}^{~(\alpha}g_{\dot{\sigma}}^{~\beta}g_{\rho]}^{~\dot{\alpha}]}$ & $g_{(\dot{\rho}}^{~(\alpha}g_{\dot{\sigma}}^{~i}g_{\rho]}^{~j]}$ & $g_{(\dot{\rho}}^{~(\alpha}g_{\dot{\sigma}}^{~i}g_{\rho]}^{~\dot{\alpha}]}$ & $g_{(\dot{\rho}}^{~(i}g_{\dot{\sigma}}^{~j}g_{\rho]}^{~k]}$\\
	\hline
	$\dot{\rho}\ell m$ & $g_{(\dot{\rho}}^{~(\alpha}g_{\ell}^{~\beta}g_{m]}^{~i]}$ & $g_{(\dot{\rho}}^{~(\alpha}g_{\ell}^{~\beta}g_{m]}^{~\dot{\alpha}]}$ & $g_{(\dot{\rho}}^{~(\alpha}g_{\ell}^{~i}g_{m]}^{~j]}$ & $g_{(\dot{\rho}}^{~(\alpha}g_{\ell}^{~i}g_{m]}^{~\dot{\alpha}]}$ & $g_{(\dot{\rho}}^{~(i}g_{\ell}^{~j}g_{m]}^{~k]}$\\
	\hline
	$\dot{\rho}\ell \rho$ & $g_{(\dot{\rho}}^{~(\alpha}g_{\ell}^{~\beta}g_{\rho]}^{~i]}$ & $g_{(\dot{\rho}}^{~(\alpha}g_{\ell}^{~\beta}g_{\rho]}^{~\dot{\alpha}]}$ & $g_{(\dot{\rho}}^{~(\alpha}g_{\ell}^{~i}g_{\rho]}^{~j]}$ & $g_{(\dot{\rho}}^{~(\alpha}g_{\ell}^{~i}g_{\rho]}^{~\dot{\alpha}]}$ & $g_{(\dot{\rho}}^{~(i}g_{\ell}^{~j}g_{\rho]}^{~k]}$\\
	\hline
	$\ell mn$ & $g_{(\ell}^{~(\alpha}g_{m}^{\beta}g_{n]}^{~i]}$ & $g_{(\ell}^{~(\alpha}g_{m}^{\beta}g_{n]}^{~\dot{\alpha}]}$ & $g_{(\ell}^{~(\alpha}g_{m}^{i}g_{n]}^{~j]}$ & $g_{(\ell}^{~(\alpha}g_{m}^{i}g_{n]}^{~\dot{\alpha}]}$ & $g_{(\ell}^{~(i}g_{m}^{j}g_{n]}^{~k]}$\\
	\hline
	$\dot{\rho}\rho\sigma$ & $g_{(\dot{\rho}}^{~(\alpha}g_{\rho}^{\beta}g_{\sigma]}^{~i]}$ & $g_{(\dot{\rho}}^{~(\alpha}g_{\rho}^{\beta}g_{\sigma]}^{~\dot{\alpha}]}$ & $g_{(\dot{\rho}}^{~(\alpha}g_{\rho}^{i}g_{\sigma]}^{~j]}$ & 0 & 0\\
	\hline
	$\ell m \rho$ & $g_{(\ell}^{~(\alpha}g_{m}^{\beta}g_{\rho]}^{~i]}$ & $g_{\ell}^{~(\alpha}g_{m}^{\beta}g_{\rho]}^{~\dot{\alpha}]}$ & $g_{\ell}^{~(\alpha}g_{m}^{i}g_{\rho]}^{~j]}$ & 0 & 0\\
	\hline
	$\ell\rho\sigma$ & $g_{(\ell}^{~(\alpha}g_{\rho}^{\beta}g_{\sigma]}^{~i]}$ & 0 & 0 & 0 & 0\\
	\hline
	\end{tabular}
	
\vskip.2in	

	\begin{tabular}{| l || c | c | c |}
	\hline
	& $\alpha\dot{\alpha}\dot{\beta}$ & $\dot{\alpha} ij$ & $\dot{\alpha}\dot{\beta}i$\\
	\hline
	$\dot{\rho}\dot{\sigma}\ell$ & $g_{(\dot{\rho}}^{~(\alpha}g_{\dot{\sigma}}^{~\dot{\alpha}}g_{\ell]}^{~\dot{\beta}]}$ & $g_{(\dot{\rho}}^{~(\dot{\alpha}}g_{\dot{\sigma}}^{~i}g_{\ell]}^{~j]}$ & $g_{(\dot{\rho}}^{~(\dot{\alpha}}g_{\dot{\sigma}}^{~\dot{\beta}}g_{\ell]}^{~i]}$\\
	\hline
	$\dot{\rho}\dot{\sigma}\rho$ & $g_{(\dot{\rho}}^{~(\alpha}g_{\dot{\sigma}}^{~\dot{\alpha}}g_{\rho]}^{~\dot{\beta}]}$ & $g_{(\dot{\rho}}^{~(\dot{\alpha}}g_{\dot{\sigma}}^{~i}g_{\rho]}^{~j]}$ & 0\\
	\hline
	$\dot{\rho}\ell m$ & $g_{(\dot{\rho}}^{~(\alpha}g_{\ell}^{~\dot{\alpha}}g_{m]}^{~\dot{\beta}]}$ & $g_{(\dot{\rho}}^{~(\dot{\alpha}}g_{\ell}^{~i}g_{m]}^{~j]}$ & 0\\
	\hline
	$\dot{\rho}\ell \rho$ & 0 & 0 & 0\\
	\hline
	$\ell mn$ & 0 & 0 & 0\\
	\hline
	$\dot{\rho}\rho\sigma$ & 0 & 0 & 0\\
	\hline
	$\ell m \rho$ & 0 & 0 & 0\\
	\hline
	$\ell\rho\sigma$ & 0 & 0 & 0\\
	\hline
	\end{tabular}
	
\vskip.2in	

	\noindent 0 means it is negative scale dimension constraint, therefore no additional constraints.

\section*{Acknowledgments}

This work is supported in part by National Science Foundation Grant No. PHY-0969739.

	\bibliographystyle{utphys}
	\bibliography{references}

\end{document}